\documentclass[12pt]{iopart}
\usepackage{graphicx}

\begin{document}

\title[Luminescence lineshape of NV centres]
{First-principles theory of the luminescence lineshape for the triplet transition in diamond NV centres}

\author{Audrius Alkauskas$^1$, Bob B. Buckley$^2$, David D. Awschalom$^{2,3}$, Chris G. Van de Walle$^1$}

\address{$^1$Materials Department, University of California, Santa Barbara, California, USA}
\address{$^2$Center for Spintronics and Quantum Computation, University of California, Santa Barbara, California, USA}
\address{$^3$Institute for Molecular Engineering, University of Chicago, Chicago, Illinois, USA}


\date{\today}

\begin{abstract}
In this work we present theoretical calculations and analysis of the vibronic structure
of the spin-triplet optical transition in diamond nitrogen-vacancy centres. The electronic
structure of the defect is described using accurate first-principles methods based on hybrid
functionals. We devise a computational methodology to determine the coupling between
electrons and phonons during an optical transition in the dilute limit. As a result, our
approach yields a smooth spectral function of electron-phonon coupling and includes
both quasi-localized and bulk phonons on equal footings. The luminescence lineshape is
determined via the generating function approach. We obtain a highly accurate description
of the luminescence band, including all key parameters such as the  Huang-Rhys
factor, the Debye-Waller factor, and the frequency of the dominant phonon mode. More
importantly, our work provides insight into the vibrational structure of nitrogen vacancy centres,
in particular the role of local modes and vibrational resonances. In particular, we find
that the pronounced mode at 65 meV is a vibrational resonance, and we quantify
localization properties of this mode. These excellent results for the benchmark diamond
nitrogen-vacancy centre provide confidence that the procedure can be applied to other defects,
including alternative systems that are being considered for applications in quantum information processing.
\end{abstract}


\submitto{\NJP}
\maketitle


\section{Introduction}

In the past decade, the negatively charged nitrogen-vacancy (NV) centre in diamond \cite{Doherty_RPP_2013}
has emerged as a very versatile solid-state system for studies of quantum information \cite{Toyli_MRS_2013}.
The main characteristics that make it unique \cite{Doherty_RPP_2013} are its paramagnetic ground state
\cite{Loubser_RPP_1978}, bright luminescence, extremely long spin coherence times \cite{Balasubramanian_NM_2009},
coupling to nearby nuclear spins \cite{Childress_Science_2006}, and the ability to initialize and read out the spin
using optical techniques \cite{Jelezko_PRL_2004a,Jelezko_PRL_2004b}. Increasingly, NV centres in bulk crystals and
nanodiamonds are used for metrological applications at the nanoscale, i.e., for measuring local magnetic \cite{Taylor_NP_2008}
and electric \cite{Dolde_Nature_2011} fields, temperature \cite{Acosta_PRL_2010,Toyli_PRX_2012,Toyli_PNAS_2013}, and pressure
\cite{Doherty_PRL_2014}.

\begin{figure}
\begin{center}
\includegraphics[width=15.5cm]{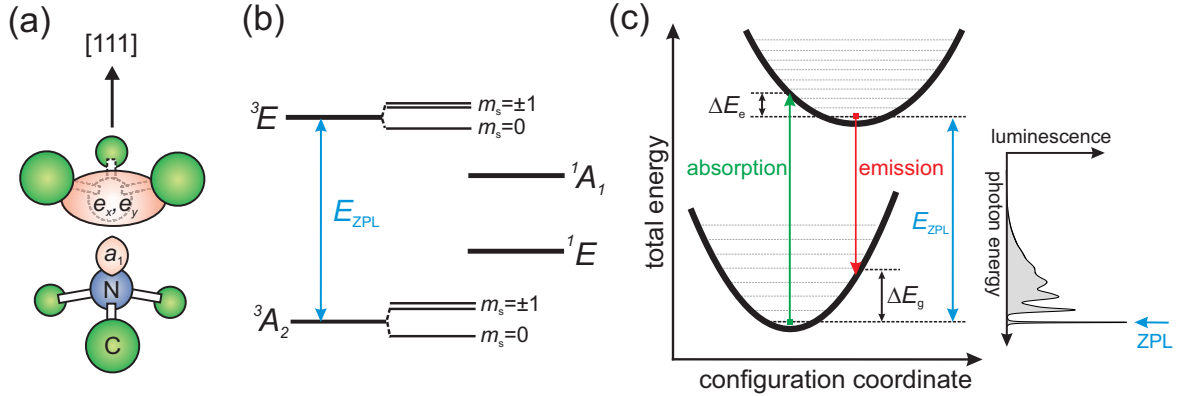}
\caption{
(a) Schematic representation of the NV centre. Green: carbon atoms; blue: nitrogen atom.
The dashed circle indicates the position of the carbon vacancy.
The pink isosurfaces depict the wavefunctions of the single-electron $a_1$ and $e$ states.
(b) Energy-level diagram of the negatively charged NV centre. The luminescence occurs
between triplet states $^3\hspace{-0.5mm}E$ and $^3\hspace{-0.5mm}A_2$. $E_{ZPL}$ is the energy of
the zero-phonon line. (c) One-dimensional cc diagram illustrating luminescence.
$\Delta E_{\{e,g\}}$ are relaxation energies in the excited and the ground state.
}
\label{fig-NV}
\end{center}
\end{figure}

The negatively charged NV centre possesses $C_{3v}$ symmetry and consists of a substitutional nitrogen atom
adjacent to a nearby carbon vacancy (figure \ \ref{fig-NV}(a)) with an additional trapped electron, the total electric
charge thus being $-1$. The electronic structure of the ground and the lowest excited states of the centre
is mainly determined by four electrons in atomically localized states of $a_1$ and $e$ ($e_{x}$ and $e_{y}$)
symmetries; the energy level diagram of the many-electron system is shown
in figure \ref{fig-NV}(b) \cite{Doherty_NJP_2011,Maze_NJP_2011}. The basics of NV physics is understood in terms
of the ground-state triplet $^3\hspace{-0.5mm}A_2$ state (configuration $a_1^2e^2$), the excited state
triplet $^3\hspace{-0.5mm}E$ state (configuration $a_1^1e^3$), and two singlet ``dark'' states $^1E$
and $^1A_1$ (configuration $a_1^2e^2$). The singlets play a crucial role in both initialization and
read-out of the ground-state spin \cite{Doherty_RPP_2013}.

Nearly all of the applications of NV centres rely on measuring photoluminescence between
$^3\hspace{-0.5mm}E$ and $^3\hspace{-0.5mm}A_2$ electronic states as a function of other
experimental parameters \cite{Doherty_RPP_2013}. At low temperatures the luminescence band \cite{Davies_PRS_1976}
consists of a sharp zero-phonon line (ZPL) at $E_{ZPL}=1.945$ eV (637 nm), and $\sim 4$ increasingly broad phonon
replicas with a phonon energy of $\sim63$-$65$ meV, as schematically shown in figure \ref{fig-NV}(c). A detailed
analysis of the experimental phonon sideband was performed by Davies \cite{Davies_RPP_1981}. In particular,
he determined the weight of the ZPL $w_{ZPL}\approx2.4$ \% and a Huang-Rhys (HR) factor
\cite{Huang_PRCL_1950}, in essence the average number of phonons emitted during the optical
transition (see Section \ref{theory} for a quantitative definition), $S=3.73$.

The relevance of the NV centre to a variety of applications and the crucial importance
of the luminescence band in all these applications raises a question: can the luminescence
lineshape, i.e., the electron-phonon coupling during the optical transition, be calculated using
first-principles calculations that require no experimental input? Such calculations should
address an accurate determination of the Huang-Rhys factor, frequencies of dominant phonon
modes, as well as the fine structure of the phonon sideband, including the coupling to long-range
acoustic phonons. Previous work \cite{Gali_NJP_2011,Zhang_PRB_2011}
has addressed the vibrational structure of NV centres to some extent; however, because
of finite-size effects the results of these calculations are somewhat ambiguous.
No first-principles calculation of the luminescence lineshape has been performed to date. Such a
calculation would provide valuable information about electron-phonon coupling at NV centres, which at present
is incompletely understood \cite{Doherty_RPP_2013}. In addition, if the theory is \emph{predictive}, it can
be applied to other defects, for example alternative systems that are currently being actively considered for
quantum information and metrology applications \cite{Weber_PNAS_2010,Koehl_Nature_2011,Gali_pss_2011,Baranov_PRB_2011},
or defects that play an important role in light-emitting diodes \cite{Reshchikov_JAP_2005,Brgoch_JPCC_2013}.

In this work we present accurate calculations of the vibronic structure pertaining to the triplet luminescence
band of the NV centre in diamond. We demonstrate that the combination of state-of-the-art first-principles methods,
in particular hybrid density functional theory \cite{Heyd_JCP_2003}, and computational techniques to address
electron-phonon coupling at large enough length scales to accurately include long-wavelength
acoustic phonons, is very successful in describing the luminescence lineshape and all the related parameters.
The experimental luminescence spectrum which serves as a benchmark for the theoretical study has been measured
in our laboratory.
The measured luminescence band is of a comparable quality to those of Refs.\
\cite{Santori_NT_2010} and \cite{Kehayias_2013}.

This paper is organized as follows. In Section \ref{theory} we outline the general theory to calculate
the vibrational structure of luminescence bands and describe our computational approach. In Section
\ref{experiment} the details of acquiring and processing the experimental spectrum are presented.
The results are presented in Section \ref{results} and analyzed in Section \ref{analysis}. Section \ref{conclusions}
contains our conclusions. The paper is supplemented with four appendices that discuss specific technical
issues in more detail.


\section{Theory and computational methodology \label{theory}}

\subsection{Luminescence \label{theory2}}

The excited state $^3\hspace{-0.5mm}E$ is an orbital doublet that forms an $E\otimes e$
Jahn-Teller system via coupling to $e$ phonon modes \cite{Davies_RPP_1981,Fu_PRL_2009,Abtew_PRL_2011}.
The Jahn-Teller effect is dynamical, since the energy splitting between the vibronic sub-levels
is larger than the barrier in the adiabatic potential energy surface $\delta \approx 10$ meV
\cite{Abtew_PRL_2011}. The presence of this effect is manifest in the broadening of the
ZPL that follows a $\sim T^5$ rather than the usual $\sim T^7$ dependence at low temperatures
\cite{Fu_PRL_2009,Abtew_PRL_2011}. However, the effect is weak \cite{Abtew_PRL_2011},
and we will neglect it when calculating the phonon sideband. One way to judge the validity of this
approximation is via an \emph{a posteriori} comparison \cite{Davies_RPP_1981}. If a linear model of
electron-phonon interactions, such as the one employed in this work, accurately describes the lineshape
of the optical transition, then the Jahn-Teller effect can be considered negligible for this particular
transition. As we show below, this turns out to be the case for the triplet luminescence at NV centres.

We also assume that the transition dipole moment $\vec{\mu}_{eg}$ between the excited and the ground
state depends weakly on lattice parameters (the Franck-Condon approximation). At $T=0$ K the absolute luminescence
intensity $I(\hbar\omega)$ (i.e., photons per unit time per unit energy) for a given photon energy $\hbar\omega$
and for one emitting centre is given by (in SI units) \cite{Stoneham}:
\begin{equation}
I(\hbar\omega)=\frac{n_{D}\omega^3}{3\varepsilon_0\pi c^3 \hbar}\left|\vec{\mu}_{eg}\right|^2
\sum_m\left| \left \langle \chi_{gm}|\chi_{e0} \right \rangle \right |^2
\delta \left(E_{ZPL}-E_{gm}-\hbar\omega\right).
\label{Lum-int}
\end{equation}
Here $n_D=2.4$ is the refractive index of diamond; $\chi_{e0}$ and $\chi_{gm}$ are
vibrational levels of the excited and the ground  state; $E_{gm}$ is the energy of the state $\chi_{gm}$,
being the sum over all vibrational modes $k$, i.e., $E_{gm}=\sum_{k}n_{k}\hbar\omega_k$; and
$n_k$ is the number of phonons of type $k$ in this state. The absolute angle-averaged value of $\mu_{eg}$
is $\sim 5.2$ Debye, as extracted from the radiative lifetime $\tau=13$ ns of the $m_s=0$ spin state of the $^3 E$
manifold \cite{Doherty_RPP_2013}. A prefactor $\omega^3$ in equation (\ref{Lum-int}) arises from the density of states
of photons that cause the spontaneous emission ($\sim\omega^2$), and the perturbing electric field of those
photons ($|\vec{\cal{E}}|^2\sim\omega$). This prefactor has to be taken into account
when determining parameters pertaining to the luminescence lineshape, and this will be discussed in Section \ref{HR3}.
Since in both the excited and the ground electronic state the system has $C_{3v}$ symmetry,
only fully symmetric $a_1$ phonons contribute to the sum in equation (\ref{Lum-int}).

The experimental determination of the absolute luminescence intensity given in equation (\ref{Lum-int}) is difficult.
Thus, in this work we will consider the normalized luminescence intensity, defined as
\begin{equation}
L(\hbar\omega)=C\omega^3\hspace{-0.5mm}A(\hbar\omega),
\label{Norm-lum-int}
\end{equation}
where
\begin{equation}
A(\hbar\omega)=
\sum_m\left| \left \langle \chi_{gm}|\chi_{e0} \right \rangle \right |^2
\delta \left(E_{ZPL}-E_{gm}-\hbar\omega\right)
\label{spectral}
\end{equation}
is the optical spectral function, and $C$ is the normalization constant: $C^{-1}=\int A(\hbar\omega) \omega^3 d(\hbar\omega$).
$I(\hbar\omega)$ is related to $L(\hbar\omega)$ via $I(\hbar\omega)=n_{D}/(3C\varepsilon_0\pi c^3 \hbar)L(\hbar\omega)$.

The evaluation of the overlap integrals $ \left \langle \chi_{gm}|\chi_{e0} \right \rangle $
immediately poses a challenge. Vibrational modes that enter into equation\ (\ref{spectral}) are not those of the
pristine bulk, but rather those of the solid with a defect.  The use of bulk modes can lead to large discrepancies
with experiment, as we will show in Section \ref{results}. Lattice imperfections induce
localized or quasi-localized vibrational modes that depend on the local electronic structure; in addition, the normal
modes in the excited state and the ground state can be in principle quite different \cite{Duschinsky_AP_1937}. This
results in highly multidimensional integrals that can in practice be evaluated only for molecules \cite{Borrelli_JCP_2003},
small atomic clusters \cite{Patrick_NC_2013}, or model defect systems \cite{Alkauskas_PRL_2012}.

Some kind of approximation is thus unavoidable.  Here we assume that (i) the normal modes that
contribute to the luminescence lineshape are still those of the solid with a defect, but (ii) the modes in
the excited electronic state are identical to those in the ground state. Such an assumption is implicit in
virtually all studies of defects in solids \cite{Markham_RMP_1959,Lax_JCP_1952}. First-principles calculations
\cite{Gali_NJP_2011,Zhang_PRB_2011},
as well as a comparison of experimental absorption and emission spectra \cite{Davies_PRS_1976}, indicate that
the assumption does not strictly hold for the NV centre. Since the more exact calculation is not feasible, the validity
of this approximation has to be checked by comparing the results with the experimental spectrum.

When vibrational modes in the ground and the excited state are identical, the optical spectral function $A(\hbar\omega)$
(equation\ (\ref{spectral})) can be calculated using a generating function approach proposed by Lax \cite{Lax_JCP_1952},
as well as Kubo and Toyozawa \cite{Kubo_PTP_1955}. The fundamental quantity that has to be calculated is the spectral function
(also called spectral density) of electron-phonon coupling \cite{Miyakawa_PRB_1970}
\begin{equation}
S(\hbar\omega)=\sum_k S_{k} \delta(\hbar\omega-\hbar\omega_k),
\label{e-ph1}
\end{equation}
where the sum is over all phonon modes $k$ with frequencies $\omega_k$, and $S_k$ is the (partial) Huang-Rhys factor
for the mode $k$. It is defined as \cite{Markham_RMP_1959}
\begin{equation}
S_k=\omega_k q_k^2 /(2\hbar)
\label{Sk}
\end{equation}
with
\begin{equation}
q_k=\sum_{\alpha i}m_{\alpha}^{1/2}(R_{e;\alpha i}-R_{g;\alpha i})\Delta r_{k;\alpha i}.
\label{qk}
\end{equation}
$\alpha$ labels atoms, $i=\{x,y,z\}$, $m_{\alpha}$ is the mass of atom $\alpha$ (carbon or nitrogen,
average atomic masses of naturally occurring isotopes were used),
$R_{\{e,g\};\alpha i}$ is the equilibrium position in the initial (excited) and the final (ground)
excited state, and $\Delta  r_{k;\alpha i}$ is a normalized vector that describes the displacement of
the atom $\alpha$ along the direction $i$ in the phonon mode $k$. One can use an alternative expression for $q_k$:
\begin{equation}
q_{k}=\frac{1}{\omega_k^2}\sum_{\alpha i}\frac{1}{m_{\alpha}^{1/2}}(F_{e; \alpha i}-F_{g; \alpha i}) \Delta r_{k;\alpha i},
\label{qk2}
\end{equation}
where $F_{e; \alpha i}-F_{g; \alpha i}$ is the change of the force on the atom $\alpha$ along the direction $i$
for a fixed position of all atoms when the electronic state of the defect changes from $^3\hspace{-0.5mm}E$ to
$^3\hspace{-0.5mm}A_2$. The latter equation directly follows from the relationship
$(\vec{R}_{e}-\vec{R}_{g})=-\hat{H}^{-1}(\vec{F}_{e}-\vec{F}_{g})$, where $\hat{H}$ is the Hessian matrix,
different from the dynamical matrix only because of additional mass prefactors in the latter.
The two formulations are completely equivalent in the harmonic approximation. In \ref{1D} we show
that if the dynamical Jahn-Teller effect is neglected the anharmonicities are indeed minute. While being
in principle equivalent, the use of equation\ (\ref{qk2}) instead of equation\ (\ref{qk}) offers a huge
advantage when dealing with large systems, i.e., when extrapolating $S(\hbar\omega)$ to the dilute limit,
and this is discussed in Section \ref{ab-initio} and \ref{supercell}.

Once $S(\hbar\omega)$ is determined, the spectral function $A(\hbar\omega)$ (equation\ (\ref{spectral})) is given as the
Fourier transform of the generating function $G(t)$ \cite{Lax_JCP_1952,Kubo_PTP_1955}:
\begin{equation}
A(E_{ZPL}-\hbar\omega)=\frac{1}{2\pi}\int_{-\infty}^{\infty}G(t)e^{i\omega t-\gamma|t|}dt.
\label{spectral2}
\end{equation}
The generating function $G(t)$ itself is defined as
\begin{equation}
G(t)=e^{S(t)-S(0)},
\label{spectral6}
\end{equation}
where
\begin{equation}
S(t)=\int_{0}^{\infty}S(\hbar\omega)e^{-i\omega t}d(\hbar\omega)
\label{spectral7}
\end{equation}
and
\begin{equation}
S\equiv	S(t=0)=\int_0^{\infty}S(\hbar\omega)d(\hbar\omega)=\sum_k S_k
\label{HR}
\end{equation}
is the total HR factor for a given optical transition. In equation\ (\ref{spectral2}) the parameter $\gamma$ represents
the broadening of the ZPL. In real situations this broadening has two contributions: the homogeneous broadening due to
anharmonic phonon interactions \cite{Silsbee_PR_1962,McCumber_JAP_1963} and the inhomogeneous broadening due to ensemble
averaging. Since neither of these two effects is modeled in our approach, $\gamma$ is chosen to reproduce the
experimental width of the ZPL.

\subsection{Huang-Rhys and Debye-Waller factors \label{HR3}}

The partial HR factor $S_k$ defined in equation\ (\ref{Sk}) is the average number of phonons of type $k$
emitted during an optical transition \cite{Huang_PRCL_1950}. The total HR factor, defined in equation\ (\ref{HR}),
is then the number of phonons off all kinds that are emitted during the same transition.
The HR factor is thus an important parameter that characterizes the vibrational
structure of the luminescence band. If (i) there was no additional prefactor
$\sim\omega^3$ in the expression for the luminescence intensity in equation\ (\ref{Norm-lum-int})
and (ii) the vibrational modes in the excited and the ground state were indeed identical,
then the weight of the zero-phonon line would be given by \cite{Davies_JPC_1974,Davies_RPP_1981,Markham_RMP_1959,Lax_JCP_1952}
$w_{ZPL} = e^{-S}.$ Since this line corresponds, by definition, to zero absorbed or emitted phonons,
$w_{ZPL}$ is often called the Debye-Waller factor, in analogy with x-ray scattering, where it represents
the ratio of the elastic to the total scattering cross section. Therefore, we also use this nomenclature
to comply with the accepted practice.

The Debye-Waller factor $w_{ZPL}$
is a quantity that is directly measurable in experiment, and its determination is therefore unambiguous.
In practical situations the HR factor is often deduced from the relationship $\widetilde {S} = -\ln (w_{ZPL})$,
where we have added a ``tilde'' to distinguish this quantity from the actual HR factor $S$, which is
defined by equation\ (\ref{HR}). $\widetilde {S}$ differs from $S$ because of the additional assumption (i).

The spectral weight in $L(\hbar\omega)$ (equation\ (\ref{Norm-lum-int})) moves to slightly higher energies
in comparison to $A(\hbar\omega)$ due to the prefactor $\omega^3$. This increases the weight of the
ZPL, $w_{ZPL}$, if determined from $L(\hbar\omega)$, and thus decreases the value of $\widetilde S$
with respect to $S$. This distinction has to be borne in mind when comparing different experimental
papers. In this paper we will consistently use $w_{ZPL}$ and $S$ in their original definitions.

\subsection{First-principles approach \label{ab-initio}}

In this work the spectral function of electron-phonon coupling $S(\hbar\omega)$ (equations\
(\ref{e-ph1})-(\ref{qk})) is calculated within density functional theory (DFT).  The electronic,
atomic, and vibrational properties of the NV centre are calculated in the supercell approach
\cite{Freysoldt_RMP_2014}, whereby one defect is embedded in a sufficiently large piece of host
material, which is periodically repeated. We take a conventional cubic cell with 8 carbon atoms
as the building block for larger supercells. The cubic supercell $N$$\times$$N$$\times$$N$,
for example, contains $M$=$8N^3$ atomic sites.

To study the electronic and vibrational structure of defects we use two different exchange-correlation
(XC) functionals: the generalized gradient approximation (GGA) in the form proposed by Perdew, Burke, and
Ernzerhof (PBE) \cite{Perdew_PRL_1996} and the screened hybrid functional of Heyd, Scuseria, and Ernzerhof
(HSE) \cite{Heyd_JCP_2003}. PBE is known to describe structural properties of many materials with high
accuracy, but the calculated band gaps of semiconductors and insulators agree poorly with experiment,
and this also affects the position of defect levels within the band gap. The HSE functional overcomes
this problem by incorporating a fraction $a=1/4$ of screened Fock exchange (screening parameter $\omega=0.2$ \AA$^{-1}$).
HSE calculations yield excellent results for excitation energies for the spin-triplet optical transition
in NV centres \cite{Gali_PRL_2009}.

The properties of the excited state $^3\hspace{-0.5mm}E$ have been calculated using the constrained
occupation method of Slater \cite{Slater}, as first applied to the NV centre by Gali \emph{et al.}
\cite{Gali_PRL_2009}. In this method one electron from the $a_1$ orbital is promoted to one of the $e$
orbitals. The electronic and the atomic structure is optimized with a hole in the $a_1$ state.
To circumvent the problems with the Jahn-Teller distortion in the excited state, resulting from
the degeneracy of the nominal $a_1 e_x^2 e_y^1$ and $a_1 e_x^1 e_y^2$ configurations, the coordinate
dependence of the total energy in the excited state is studied here by constraining the
configuration to $a_1 e_x^{1.5} e_y^{1.5}$. This is a practical solution to restrict the excited state
density to be the average of the two degenerate configurations, retaining a $C_{3v}$ symmetry.

We find that while the integrated parameters, for example the total HR factor $S$ (equation\ (\ref{HR}))
converge quickly when the size of the defect supercell is increased, the convergence
of the spectral function $S(\hbar\omega)$ (equation\ (\ref{e-ph1})) is significantly slower.
This is a particular concern for the spectral function at lower energies, i.e., coupling
to long-range acoustic phonons. As an example, let us consider a $2$$\times$$2$$\times$$2$ simple cubic
supercell, containing 64 lattice sites. Without a defect, the lowest energy $\Gamma$-point vibration of such a
supercell corresponds to the bulk transverse acoustic (TA) mode at the $\Lambda$ point with an energy of
about $68$ meV. This is even higher than the experimentally determined energy of the dominant phonon mode at
the NV centre, being $\hbar\omega_0$=$63$-$65$ meV. Clearly, supercells of this size are insufficient to
determine $S(\hbar\omega)$.

To obtain converged results and determine the nature of vibrational states, we have performed
calculations for a series of supercells: from $2$$\times$$2$$\times$$2$  (64 sites)
up to $11$$\times$$11$$\times11$  (10648 sites). Since a direct approach for supercells
containing more than a few hundred atoms is computationally too demanding,
we have developed a special methodology to achieve this goal. First, in \ref{1D} we show that it
is an excellent approximation to calculate vibrational properties at the PBE level, since the relevant
vibrational modes are very similar in the PBE as compared to the HSE functional. This presents huge computational
savings, since HSE calculations are up to two orders of magnitude more expensive. Then in \ref{supercell}
we present a methodology to calculate vibrational spectra and spectral functions $S(\hbar\omega)$ for very
large systems. In short, the procedure is as follows. Partial Huang-Rhys factors for large systems are
determined from equation (\ref{qk2}). Forces $\vec{F}_{\{e,g\}}$ in the large supercell $N$$\times$$N$$\times$$N$
($N>3$), needed in that approach, are obtained from the calculation of a smaller supercell ($4$$\times$$4$$\times$$4$)
via a suitable embedding procedure, explained in \ref{supercell}. For these large defect supercells, vibrational
modes and frequencies, that also appear in expression (\ref{qk2}), have been determined by diagonalizing the
dynamical matrix constructed from dynamical matrices of bulk diamond and NV centre in the $3$$\times$$3$$\times$$3$
supercell. The validity of the procedure relies on the fact that the dynamical matrix of diamond is rather
short-ranged. Specific parameters of the procedure are determined from accurate convergence tests, and are
discussed in \ref{supercell}.

Defect calculations have been performed with the {\sc vasp} code \cite{VASP1,VASP2}, and the interaction
with ionic cores was described via the projector-augmented wave (PAW) formalism \cite{Bloechl_PRB_1994}.
A kinetic energy cutoff of 400 eV (29.4 Ry) has been used for the expansion of electronic wavefunctions.
For the $2\times 2 \times 2$ supercell the Brillouin zone was sampled using a $2\times2\times2$ $k$-point
mesh, and $\Gamma$-point sampling was used for larger supercells.

To produce additional insights, we have also calculated $S(\hbar\omega)$ (equation\ (\ref{Sk})) and $L(\hbar\omega)$
(equation\ (\ref{Norm-lum-int})) with an additional assumption, namely that phonon modes that contribute to the
luminescence lineshape are those of the unperturbed host \cite{Kretov_JL_2012,Kretov_RJPCA_2013}. For this purpose we
have determined the vibrational modes of bulk diamond using density functional perturbation theory \cite{Baroni_RMP_2001},
reproducing earlier calculations \cite{Pavone_PRB_1993}. Vibrational modes were determined on a very fine $27\times27\times27$
$k$-point grid close to the Brillouin zone centre, and a courser $9\times9\times9$ grid elsewhere. These calculations
have been performed using the {\sc quantum espresso} code \cite{QuantumEspresso} within the local density approximation
\cite{Perdew_PRB_1981}; this XC functional describes phonons modes of \emph{bulk} diamond very well \cite{Pavone_PRB_1993}.
To evaluate $S(\hbar\omega)$, the modes were mapped  to the $\Gamma$-point of the desired supercell. The contributions
from the vacancy site are set to zero in equations\ (\ref{qk}) and equation\ (\ref{qk2}), while the mass of the nitrogen
atom was set to be equal to that of the carbon atom
in this case.


\section{Experimental spectrum \label{experiment}}

The NV centre photoluminescence (PL) spectrum was taken at 8 K on an ensemble of NV centres using a home-built
confocal setup.  The diamond sample used was a {\sc Sumitomo} high-pressure, high-temperature grown Ib
{\sc Sumicrystal} with a specified nitrogen content of $30-100$ parts per million.  The sample was irradiated with
2 MeV electrons at a dose of $1\times10^{17}$ electrons/cm$^2$ and annealed at 850 $^{0}$C for 2 hours to generate a
high density of NV centres within the bulk.  The NV centres were photo-excited with 532 nm light with sufficiently low
intensity to suppress the luminescence of neutral (NV$^0$) centres.  Subsequent PL was collected into a spectrometer
with $\sim$0.3 meV spectral resolution.  The spectrum intensity was calibrated by measuring the nominally-known spectrum
of an {\sc OceanOptics} LS-1-LL tungsten halogen light source placed at the same position
of the diamond sample within the optical setup.

The experimentally obtained spectrum was normalized to 1 for comparison with the theoretical calculations.
For normalization purposes the low-energy tail of the spectrum was modeled as an exponential function.
The weight of the zero-phonon line (Debye-Waller factor) was determined to be $\sim3.2$\%.
This corresponds to $\tilde{S}=3.45$, in very close agreement with Ref.\ \cite{Kehayias_2013}.
The actual Huang-Rhys factor can be estimated to be $S\approx 3.85\pm0.05$.

\section{Results \label{results}}

\subsection{Excitation energies \label{prelim}}

For the $4$$\times$$4$$\times$$4$ supercell, the largest system for which we have performed
actual electronic structure calculations, $E_{ZPL}$ was calculated to be 1.757 eV using the PBE functional,
and 2.035 eV using the HSE functional. The latter is thus much closer to the experimental value of 1.945 eV.
Our calculations agree with those of Gali \emph{et al.}\ \cite{Gali_PRL_2009} and
Weber \emph{et al.} \cite{Weber_PNAS_2010}. The HSE functional is clearly superior for describing the local
electronic structure of the NV centre \cite{Gali_PRL_2009}. The difference of about 0.1 eV between the experimental
and calculated ZPL is within the error bar of the HSE calculations, but would complicate direct comparisons between
theoretical and experimental lineshapes. To enable a more meaningful comparison, in all subsequent analysis we set
$E_{ZPL}$ to the experimental value. Thus, the broadening of the ZPL $\gamma$ in equation (\ref{spectral2})
and the value $E_{ZPL}$ are the sole instances where information from experiment has been used in the
theoretical results.

\subsection{Spectral function of electron-phonon coupling $S(\hbar\omega)$ \label{spectral5}}

We first analyze the convergence of $S(\hbar\omega)$ when the size of the supercell is increased.
In addition to providing justification for the computational procedure, such a study provides
insights into the origin of vibrational modes that contribute to the phonon sideband.

\begin{figure}
\begin{center}
\includegraphics[width=12cm]{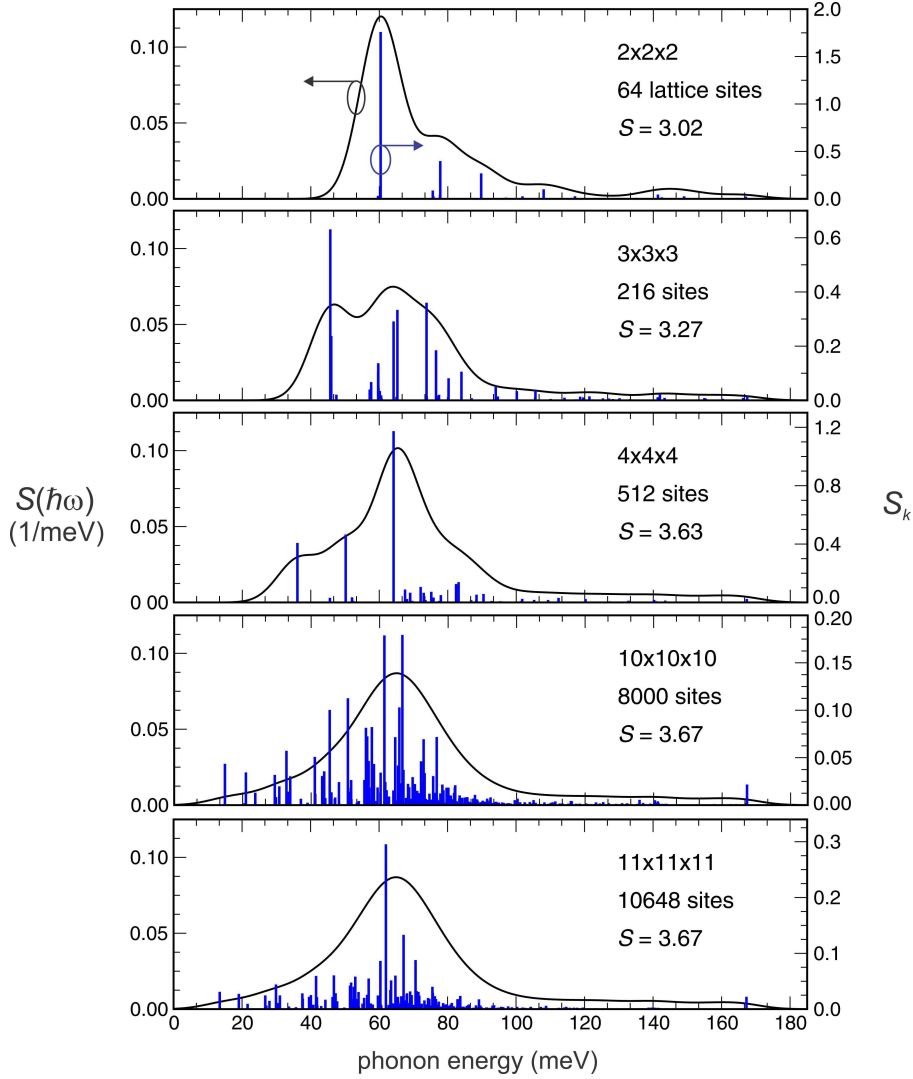}
\caption{Spectral functions $S(\hbar\omega)$ (equation\ (\ref{e-ph1})) and partial Huang-Rhys factors (equation\ (\ref{Sk}))
pertaining to the spin-triplet optical transition at NV centres for increasingly larger supercells,
from $2$$\times$$2\times$$2$ to $11\times$$11\times$$11$ (some intermediate results are not shown).
$S(\hbar\omega)$: left vertical axes and black solid lines; to enable a meaningful comparison
the range of these vertical axes is the same for all supercells. $S_k$: right vertical axes and blue bars;
the range of these vertical axes decreases for larger supercells.
}
\label{Sw-conv}
\end{center}
\end{figure}

In figure \ref{Sw-conv} we show $S(\hbar\omega)$ (equation\ (\ref{e-ph1})) and partial
Huang-Rhys factors (equation\ (\ref{Sk})) as a function of the supercell size,
from $2\times$$2\times$$2$ to $11\times$$11\times$$11$ (results for five intermediate supercells are omitted).
The range of the left vertical axes for $S(\hbar\omega)$ was kept identical for all supercells, but
note that this is not the case for the right vertical axes that apply to $S_k$. For the calculation of $S(\hbar\omega)$
$\delta$-functions in equation\ (\ref{e-ph1}) were replaced by Gaussians with widths $\sigma=6$ meV. HSE
results are discussed here.

In the case of the smallest $2\times2\times2$ supercell, only a few phonon modes contribute to $S(\hbar\omega)$.
The most important of these are modes with energies 60.4 meV and 77.8 meV, in complete agreement with the results of
Gali \emph{et al.} \cite{Gali_NJP_2011} (59.7 and 77.0 meV), who studied local vibrational modes for this size of
supercell. While the energy of the first mode is close to the energy of the most pronounced phonon mode seen in
experiment, this agreement is largely fortuitous, since, as mentioned in Section \ref{ab-initio}, the lowest-energy
bulk TA phonon mode in this supercell has a similar energy. The low-energy tail that represents the coupling to
long-range phonons ($<$60 meV) is completely missing for this supercell, and the total HR factor $S=3.02$
(inset of figure \ref{Sw-conv}) is 20\% smaller than the converged value $S=3.67$.

In the case of a larger $3\times3\times3$ supercell, the most dominant vibration is the 45.9 meV mode. This result
is an artifact resulting from the use of a small cell, since this vibration corresponds to the lowest-energy
$\Gamma$-point bulk TA phonon---it is not an actual defect-derived mode. While the total HR factor increases
to $3.27$, $S(\hbar\omega)$ is still far from converged. This emphasizes possible dangers in drawing
conclusions about local vibrational modes from small-size supercells \cite{Zhang_PRB_2011}.

\begin{figure}
\begin{center}
\includegraphics[width=10cm]{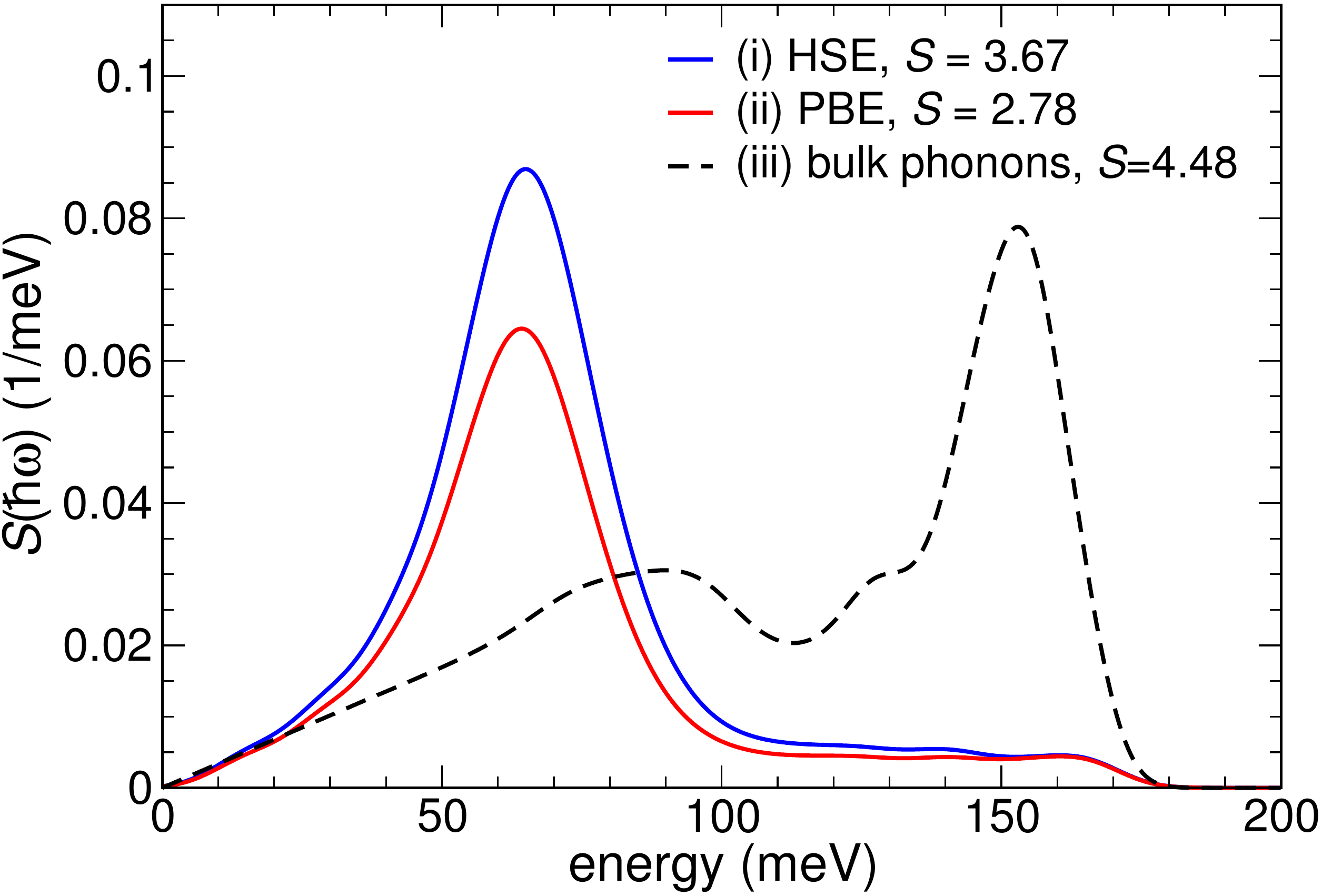}
\caption{Comparison of $S(\hbar\omega)$ calculated using three theoretical approaches: (i) the hybrid functional HSE,
(ii) the GGA functional PBE, and (iii) by using bulk phonons, as explained in the text.
}
\label{Sw-all}
\end{center}
\end{figure}

When increasing the size of the supercell further, $S(\hbar\omega)$ slowly attains its converged form.
figure \ref{Sw-all} shows that the spectral function is essentially converged for the two largest supercells
we use, even though there are still apparent changes in individual partial HR factors $S_k$.
The peak of $S(\hbar\omega)$ occurs at $\hbar\omega_0=65$ meV, in excellent agreement with experimental findings (see Sec.\
\ref{comp-lumin}). This is the first time that theoretical calculations yield the energy of the peak decisively.
Interestingly, the \emph{total} HR factor, i.e. the integral of $S(\hbar\omega)$, is within $\sim$1\% of the
converged value already for the $4\times4\times4$ supercell.

Figure\ \ref{Sw-conv} also allows us to draw the following conclusions about lattice distortions or, equivalently,
coupling to phonons, that occur during the $^3\hspace{-0.5mm}E \rightarrow\hspace{0.5mm}^3\hspace{-1mm}A_2$ optical
transition:

\begin{enumerate}
\item The $65$ meV vibration is not a localized phonon mode, but a defect-induced vibrational resonance:
it occurs within the spectrum of bulk phonon modes (0-167 meV). In figure \ref{Sw-conv} this result is
evident from the fact that for larger supercells this mode splits into many closely spaced modes, with a
simultaneous decrease of their absolute contributions. The 65 meV resonance is induced by the NV centre itself,
and cannot be understood solely by considering bulk phonon spectrum. This is demonstrated in Figure \ref{Sw-all}
and discussed in more detail in section \ref{comp-lumin}.

\item In agreement with a general theory of vibrational broadening of luminescence lines \cite{Silsbee_PR_1962},
the spectral function is linear for small energies, i.e., $S(\hbar\omega)=\alpha\hbar\omega$ for $\omega\rightarrow0$.
Indeed, partial HR factors corresponding to acoustic modes scale like $1/\omega$, which, multiplied
with the density of states of acoustic modes $\sim\omega^2$, yields this linear dependence. While this
general behaviour is known \cite{Silsbee_PR_1962}, we emphasize that the {\it prefactor} to the linear dependence is
system dependent, and only accurate atomistic calculations such as the ones presented here can provide the actual
value. In our case we obtain $\alpha\approx3.6\times10^{-4}$ meV$^{-2}=360$ eV$^{-2}$. Interaction via
acoustic phonons has been recently proposed as a promising mechanism to couple two NV centres
in nanodiamonds \cite{Albrecht_NJP_2013}. The coupling of isolated qubits is essential for
any quantum computing protocol. Our calculations provide information about the coupling of NV
centers to acoustic phonons in bulk diamond, and can be useful pursuing the ideas proposed in Ref.\
\cite{Albrecht_NJP_2013} ideas further.

\item 99\% of the lattice distortions due to the optical transition, as quantified by their contribution
to $S(\hbar\omega)$, occur within $\sim 12$ \AA$ $ of the NV centre. This follows from our finding that the
total HR factor for the $4$$\times$$4\times$$4$ supercell is within 1\% of the converged value. However, long-range
relaxations, while contributing little to the total HR factor $S$, are manifest in the low-frequency
part of $S(\hbar\omega)$, and are actually observed in the luminescence lineshape (see Section \ref{comp-lumin}).

\end{enumerate}

In figure \ref{Sw-all} we show a comparison of $S(\hbar\omega)$ calculated using three different approaches.
From here on we use the following notation when we refer to our calculations: (i) ``HSE'' refers to calculations
where atomic displacements or forces in equations\ (\ref{qk}) and (\ref{qk2}) are calculated using the HSE hybrid
functional, but vibrational modes are calculated using the PBE functional. As discussed in Section \ref{ab-initio}
and \ref{1D}, calculations for smaller supercells show that vibrational modes calculated at the PBE level are
very similar to HSE results. (ii) ``PBE'' refers to calculations in which all quantities are determined at the PBE level.
In both (i) and (ii), vibrational modes correspond to the defect system. (iii) ``Bulk phonons'' refers to calculations in which
atomic distortions or forces were determined at the HSE level, as in (i), but vibrational modes correspond to those of
the unperturbed host. The comparison of (i) and (iii) should inform us whether the introduction of the defect modifies
the vibrational spectrum, and whether the phonon sideband can be understood by considering bulk modes alone.

$S(\hbar\omega)$, calculated at the PBE level, is qualitatively very similar to the HSE result. The function has a peak
at $\hbar\omega=64$ meV, but the absolute value of $S(\hbar\omega)$ is smaller for almost all energies.
In particular, the total HR factor is $S=2.78$, a quarter smaller than in HSE. In contrast, when
the bulk phonon spectrum is used, $S(\hbar\omega)$ is even qualitatively completely different. In this case
the spectral function closely follows the density of vibrational states of bulk diamond
\cite{Pavone_PRB_1993,Zaitsev_PRB_2000}, with a pronounced peak at $\hbar\omega\approx$150 meV.
The total HR factor is 4.48 in this case. However, the coupling to low-energy ($<$20 meV) acoustic modes
is very similar to cases (i) and (ii); indeed long-range phonons are expected to be little affected by the presence
of the defect.

\subsection{Comparison with experiment: luminescence lineshape and Huang-Rhys factors}
\label{comp-lumin}

In figure \ref{spectra} we compare the luminescence lineshape $L(\hbar\omega)$ (equations
(\ref{Norm-lum-int}) and (\ref{spectral2})), calculated using the HSE functional,
with the experimental one. The agreement between theory and experiment  is extremely good.
Not only is the overall shape of the luminescence band described correctly, but all the specific features are described very accurately.
In particular:

\begin{figure}
\begin{center}
\includegraphics[width=10cm]{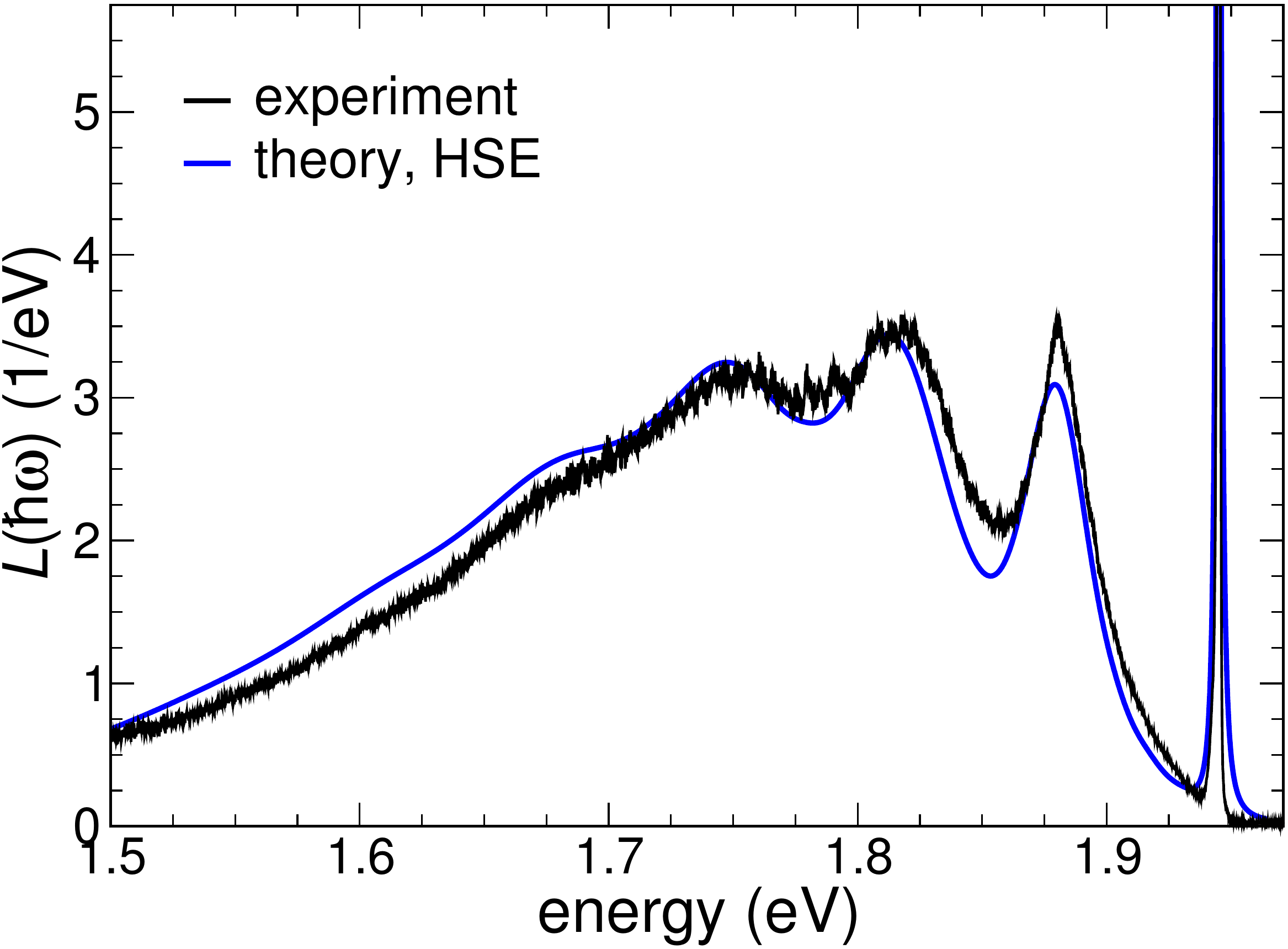}
\caption{Measured (black solid line) and calculated (blue solid line) normalized luminescence lineshapes for NV
centers in diamond in the energy range 1.5$-$2.0 eV. The calculations have been performed at the HSE level, as explained in the text.
The low-energy tail ($<$1.5 eV) of the experimental spectrum is less reliable because of calibration issues.
}
\label{spectra}
\end{center}
\end{figure}

\begin{enumerate}
\item The weight of the ZPL of the theoretical spectrum $w_{ZPL}=3.8$\% is
very close to the experimental result $w_{ZPL}=3.2$\%. Both of these quantities have been determined
directly from luminescence lineshapes shown in figure \ref{spectra}, as discussed in Section \ref{HR3}.
The theoretical Huang-Rhys factor $S=3.67$ is thus also very close to the experimental HR factor
$S=3.85\pm0.05$; the latter has been extracted from the experimental spectrum as described in Sec.\ \ref{HR3}.

\item
Both the experimental and the theoretical band show about 4 increasingly broad phonon replicas.
The theoretical phonon frequency $\hbar\omega_0$=$65$ meV is in very good agreement
with the experimental value $\hbar\omega_0$=$64$ meV.

\item
The fine structure near the ZPL, which is representative of the coupling to acoustic phonons, agrees closely.

\end{enumerate}

We conclude that calculations based on hybrid density functionals describe the vibrational properties and
the luminescence lineshape of NV centres with a very high accuracy.

\begin{figure}
\begin{center}
\includegraphics[width=10.0cm]{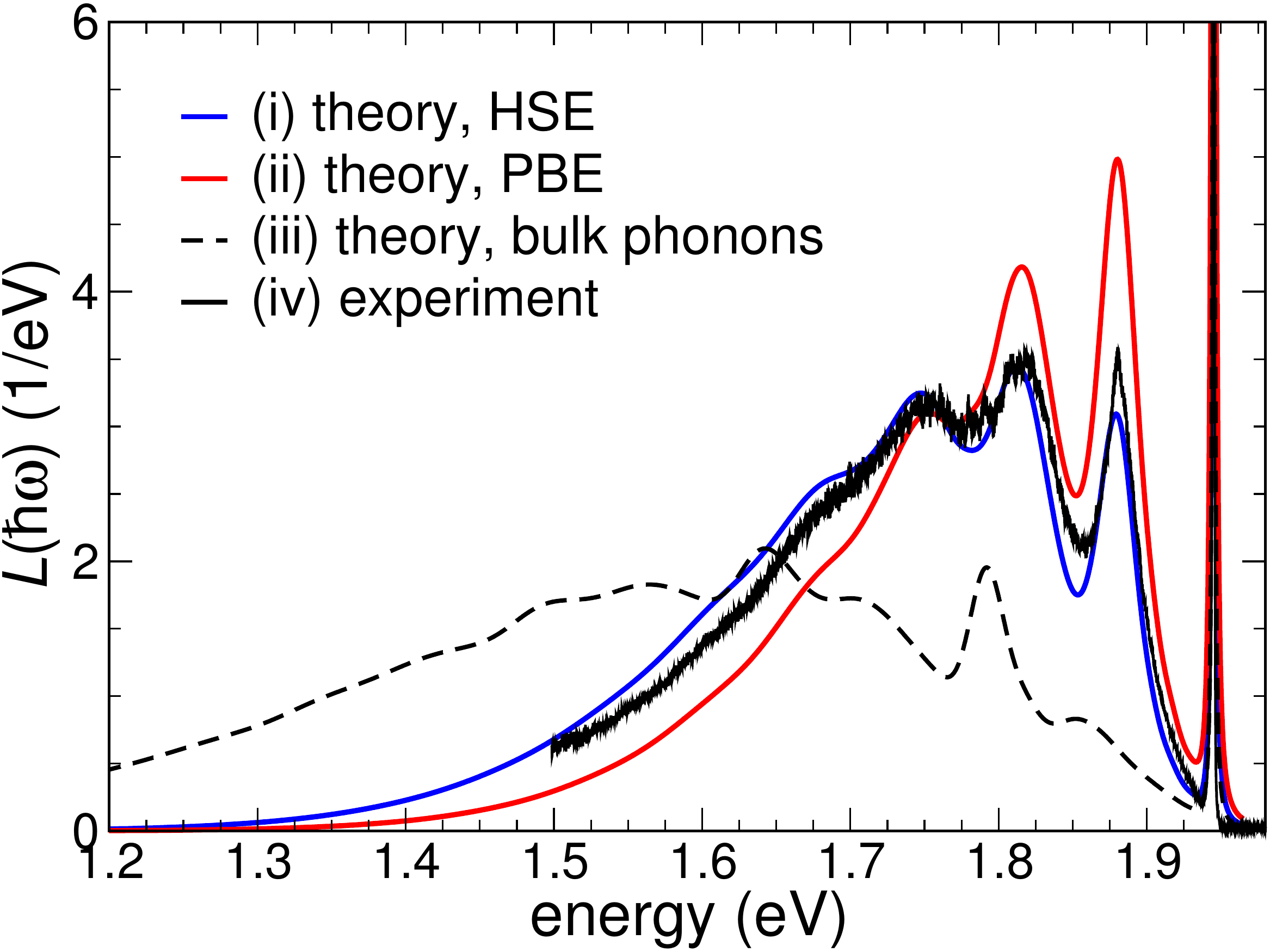}
\caption{Comparison of luminescence lineshapes calculated using the three different theoretical approaches described in
Section \ref{spectral5} (curves (i)-(iii)) with the experimental one (curve (iv)). (i) and (iv) are the same as in
figure \ref{spectra}.}
\label{spectra-comp}
\end{center}
\end{figure}

In figure \ref{spectra-comp} we present luminescence lineshapes calculated using all the three different theoretical
approaches discussed in Section \ref{spectral5}. The experimental curve and the one that corresponds to the HSE
functional are the same as those in figure \ref{spectra}. The lineshape calculated at the PBE level is qualitatively
similar to the HSE one, but there are quantitative differences. In particular, the weights of the first
two phonon replicas are larger, and the overall band is narrower. Figure~\ref{spectra-comp} also shows that when
bulk phonons are used instead, the calculated luminescence lineshape bears no resemblance to
the experimental curve: it is much broader and has a very different fine structure. This result clearly shows that
the consideration of the bulk phonon spectrum is not sufficient to understand the phonon sideband, challenging the
discussion of Ref.\ \cite{Kehayias_2013}. Taking into account vibrational modes of the defect system is essential.

\section{Analysis: localized vs. delocalized phonon modes \label{analysis}}

In Section \ref{spectral5} we mentioned that the $65$ meV phonon that
dominates the phonon sideband, is not a localized mode, but rather a vibrational resonance.
This means that it is formed by a continuum of vibrational modes, all of which have a larger weight
close to the defect, but none of which are strictly localized in real space. This mode
gives rise to a peak in $S(\hbar\omega)$ with a full width at half-maximum (FWHM) of about
32 meV (figures\ \ref{Sw-conv} and \ref{Sw-all}).

\begin{figure}
\begin{center}
\includegraphics[width=14.0cm]{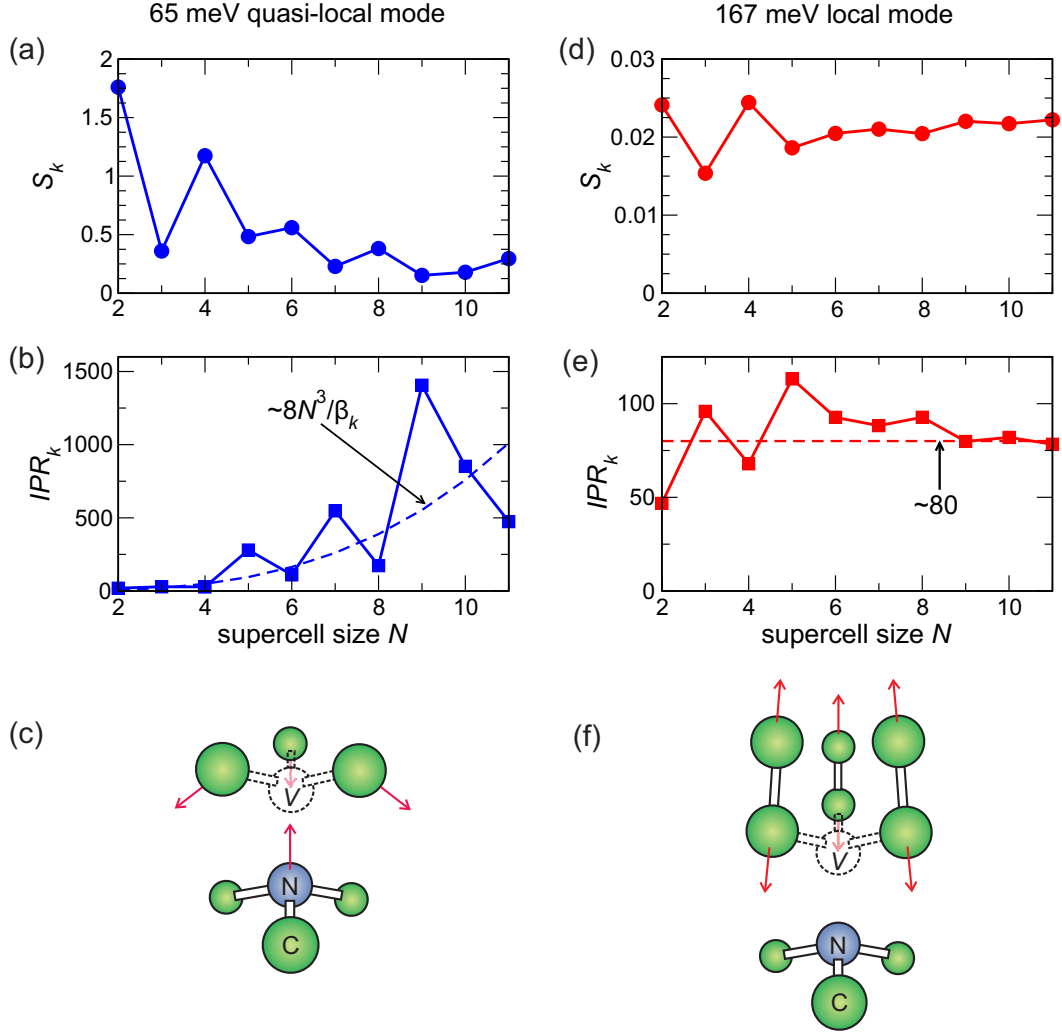}
\caption{Partial Huang-Rhys factors $S_k$ ((a) and (c), equation\ (\ref{Sk})),
inverse participation ratios $IPR_k$ ((b) and (d), equation\ (\ref{IPR})),
and vibrational patterns ((e) and (f)) for phonon modes with frequencies $\sim65$ meV
((a), (b), and (c)) and $\sim167$ meV ((d), (e), and (f)) as
a function of the supercell size.}
\label{HR-conv}
\end{center}
\end{figure}

We illustrate the fact that the 65 meV is a vibrational resonance in the following way.
For each defect supercell studied we choose the individual phonon mode that has the largest
Huang-Rhys factor in the energy range $49-81$ meV (right axis, figure \ref{Sw-conv}). The energy
range corresponds to a FWHM of the 65 meV peak in the converged function $S(\hbar\omega)$.
In figure \ref{HR-conv} (a), we plot this largest value of the partial Huang-Rhys factor $S_k$
as a function of the supercell size $N$. The value of $S_k$ for this mode decreases steadily, albeit with some
oscillations, as a function of supercell size. Since the total HR factor does not change much
as the system size grows, the decrease of this particular $S_k$ is
compensated by an increase in other phonon modes (figure \ref{Sw-conv}). This is a signature of a vibrational
resonance, which is also called a quasi-local mode.

To gain more insight, we study the inverse participation ratio (IPR) for the mode $k$ \cite{Bell_JPCC_1970}:
\begin{equation}
IPR_k = \frac{1}{\sum_\alpha p_{k;\alpha}^2},
\label{IPR}
\end{equation}
where
\begin{equation}
p_{k;\alpha} = \sum_{i}\Delta r_{k; \alpha i}^2.
\end{equation}
$IPR_k$ defined in this way measures the number of atoms onto which the vibrational mode
is localized. If, e.g., only one atom vibrates for a given mode, $IPR=1$. If all $M$ atoms
in the supercell vibrate with the same amplitude, $IPR=M$.
Note that the definition in equation\ (\ref{IPR}) is different from the one used in Ref.\ \cite{Zhang_PRB_2011}
to analyze vibrational modes of the NV centre in a 216-atom supercell, and is more in line with the traditional
definition \cite{Bell_JPCC_1970}.

In figure \ref{HR-conv}(b), the $IPR_k$ for the most pronounced mode in the energy range
$49-81$ meV is shown as a function of the supercell size $N$. For all supercell sizes
$IPR_k$ is but a fraction of the total number of atoms $M$, but steadily increases with
$N$, albeit with similar oscillations as for $S_k$. This underpins the finding that the 65 meV mode
is a vibrational resonance. This resonance represents the lion's share of the distortion of
the defect geometry (cf.\ equations (\ref{qk}) and (\ref{qk2})). It has the largest amplitude
on the four atoms surrounding the vacancy, and the vibrational pattern is shown in Fig. \ref{HR-conv}(c).
The N atom is vibrating along the defect axis, while the vibrational vectors of the C atoms
form an angle of $\sim$110$^{\circ}$ with this axis.

By analyzing partial Huang-Rhys factors and inverse participation ratios of all the modes we
were able to identify a few other, weaker resonances. These are modes with frequencies
161, 134, and 120 meV (in the order of decreasing localization). All of these weaker resonances
were recently identified in the experiment of Kehayias \emph{et al.} \cite{Kehayias_2013}.
The 153 meV resonance seen in the same experiment is not very pronounced in our calculations.
As a measure of localization we define the ``localization ratio'' $\beta$, which is the ratio
of the number of atoms in the supercell ($M=8N^3$) to the largest $IPR_k$ corresponding to
one of these resonances:
\begin{equation}
\beta_k=8N^3/(IPR_k)
\label{beta}
\end{equation}
We obtain the actual value of $\beta_k$ by fitting the $IPR_k$ for a given mode
with a function $8N^3/\beta_k$ (see figure \ref{HR-conv}(b)).
The larger the ratio $\beta_k$, the more pronounced the resonance. For a truly localized mode in the
limit $M\rightarrow\infty$, $\beta_k$ would be infinite, since for a localized mode $IPR_k$
remains constant as $M$ increases. The results are summarized in Table \ref{table_modes}. For example,
the localization ratio $\beta_k$ for the 65 meV mode is $\sim$11. Values for the ``localization ratio''
should be considered as rough estimates, but they are useful when comparing different modes.

\begin{table}
\caption{Quasi-local and local modes of $a_1$ symmetry that couple to the $^3\hspace{-0.5mm}E\rightarrow^3\hspace{-0.5mm}A_2$ optical
transition.}
\label{table_modes}	
\begin{indented}
\item[] \begin{tabular}{@{}c c }
\br
\multicolumn{2}{c}{Quasi-local modes}\\
\mr
Energy  (meV) & Localization ratio $\beta_k$ \\
\mr
65   &  $\sim$11 \\
120  &  $\sim$3  \\
134  &  $\sim$6  \\
161  &  $\sim$8  \\
\br
\multicolumn{2}{c}{Local mode}\\
\mr
Energy  (meV) & $IPR_k$ \\
\mr
167  &  $\sim$80 \\
\br
\end{tabular}
\end{indented}
\end{table}

Together with these vibrational resonances, we \emph{do} find one truly localized defect-induced
phonon mode. In figure \ref{HR-conv}(d) we show $S_k$ as a function of the supercell size for a phonon
mode with a frequency $\approx$167 meV, which is slightly ($\approx$0.2 meV) above the theoretical bulk
phonon spectrum. Increasing the size of the system, $S_k$ approaches a constant value of
$\approx$0.02. When the size of the supercell grows, the $IPR_k$ of this mode also approaches a
constant value of $\sim$80 (figure \ref{HR-conv}(e)). In analogy with shallow defect levels with energies
close to bulk band edges, one could name this mode a shallow defect-localized vibration.
While this mode is ``shallow'', half of its total weight is distributed
over 6 carbon atoms: 3 that are immediately adjacent to the vacancy,
and 3 more that are nearest neighbours of the first trio along the defect axis.
The vibrational pattern associated with this vibration is shown in figure \ref{HR-conv}(f).
It is an optical mode with vibrational vectors of all atoms only slightly off the $z$ direction (by $\sim$14$^{\circ}$)
due to the influence of the defect. The participation of the nitrogen is negligible in this vibration.

The 167 meV mode contributes less than 1 \% to the total HR factor of 3.67, and therefore its role
in the formation of the phonon sideband is not very significant. However, since this is a truly localized
vibrational mode, it can play an important role in other physical processes at nitrogen-vacancy centres.
Kehayias \emph{et al.} \cite{Kehayias_2013} recently found that a phonon mode that has the signature
of a localized vibration and an experimental energy of 169 meV plays a
noticeable role in the infrared transition $^1\hspace{-0.5mm}E\rightarrow^1\hspace{-1.4mm}A_1$.
Due to very similar atomic geometries
of the $^3\hspace{-0.5mm}A_2$, $^1\hspace{-0.5mm}A_1$ and $^1\hspace{-0.5mm}E$
electronic states \cite{Toyli_PRX_2012,Doherty_NJP_2011,Maze_NJP_2011,Kehayias_2013}
we suggest that the localized phonon mode found in our current study is the same
as the one observed in the experiments of Kehayias \emph{et al.} \cite{Kehayias_2013}.

\section{Conclusions \label{conclusions}}

In this work we have developed a first-principles methodology to calculate
the vibrational structure of defect luminescence bands. Both localized, quasi-localized,
and bulk phonons are taken into account on equal footing. The methodology was applied to study the
phonon sideband pertaining to the 1.945 eV spin-triplet transition at
nitrogen-vacancy centres in diamond. Calculations based on hybrid density functional theory yield
a luminescence lineshape and all related parameters that are in excellent agreement with experiment.
The phonon sideband is dominated by a vibrational resonance with an energy of $\sim65$ meV,
but a few other, weaker resonances, are also identified. 99 \% of all atomic relaxations that
contribute to the phonon sideband occur within $\sim$12 \AA $ $ of the defect, but the interaction with
long-range acoustic phonons is also directly manifest in the luminescence spectra close to the zero-phonon line.
We find a truly localized phonon mode slightly above the phonon spectrum of bulk diamond. While this mode,
being localized on $\sim$ 75 atoms, contributes little to the spin-triplet optical transition, it can play
an important role in other physical processes at this defect, as recent experiments suggest \cite{Kehayias_2013}.
Our findings provide a deeper understanding of the coupling of electronic states to
$a_1$ phonon states at nitrogen-vacancy centres. The success of the computational
methodology developed here provides confidence that it can be fruitfully applied to other systems of
high current interest that exhibit a complex vibrational structure of luminescence bands
\cite{Weber_PNAS_2010,Koehl_Nature_2011,Gali_pss_2011,Baranov_PRB_2011,Reshchikov_JAP_2005,Brgoch_JPCC_2013}.

\section*{Acknowledgments}
AA was supported by the Office of Science of the U.S.~Department of Energy (Grant No.~DE-SC0010689),
and by the Swiss NSF (PA00P2$\_$134127) during the initial phase of the project (2011-2012).
Experimental work at UCSB was supported by the Air Force Office of Scientific Research
and the Defense Advanced Research Projects Agency. We thank A. L. Falk and D. M. Toyli for useful comments on the
manuscript. This research used resources of the National Energy Research Scientific Computing centre, which is supported by
the Office of Science of the U.S.~Department of Energy under Contract No. DE-AC02-05CH11231,
and the Extreme Science and Engineering Discovery Environment (XSEDE),
supported by NSF (OCI-1053575 and NSF DMR07-0072N)

\appendix

\section{One-dimensional configuration coordinate diagram \label{1D}}

The methodology outlined in Section \ref{theory2} relies on the use of the harmonic approximation.
This ensures the equivalence of formulations based on equations\ (\ref{qk}) and (\ref{qk2}).
Also, in Section \ref{ab-initio} we have mentioned that the vibrational modes that are
relevant to describe the phonon sideband of the triplet luminescence are very similar
for the two functionals, PBE and HSE, used in this work. The purpose of this appendix
is to illustrate these points.

We map the potential energy surface in the ground state
and the excited state along the line in the configuration space that linearly interpolates between the equilibrium
geometries in the two states. This special mode corresponds to a displacement of an atom $\alpha$ along the
direction $i=\{x,y,z\}$ that is proportional to
$\Delta R_{\alpha i} = R_{e; \alpha i} - R_{g; \alpha i}$.
In this one-dimensional model the generalized configuration coordinate $Q$ for
values of atomic positions $R_{\alpha i}$ that correspond to this displacement is
$Q^2 = \sum_{\alpha i} m_{\alpha}\left(R_{\alpha i} - R_{g; \alpha i}\right)^2.$
The equilibrium geometry of the ground state corresponds to $Q=0$, while that of the excited states corresponds to
$Q=\Delta Q$, where
\begin{equation}
\left(\Delta Q\right)^2 = \sum_{\alpha i} m_{\alpha} \Delta R_{\alpha i}^2.
\label{vib2}
\end{equation}
A related quantity
$\left(\Delta R\right)^2 = \sum_{\alpha i} \Delta R_{\alpha i}^2$.
is also useful in analyzing theoretical results, and can be alternatively used as a measure of
atomic displacements during optical excitation.
The plot that shows the dependence of total energies in the ground and the excited states
$E_{\{e,g\}}$ as a function of $Q$ is called the configuration coordinate (cc) diagram  \cite{Stoneham}
(cf.\ figure \ref{fig-NV}(c)).

\begin{figure}
\begin{center}
\includegraphics[width=10.0cm]{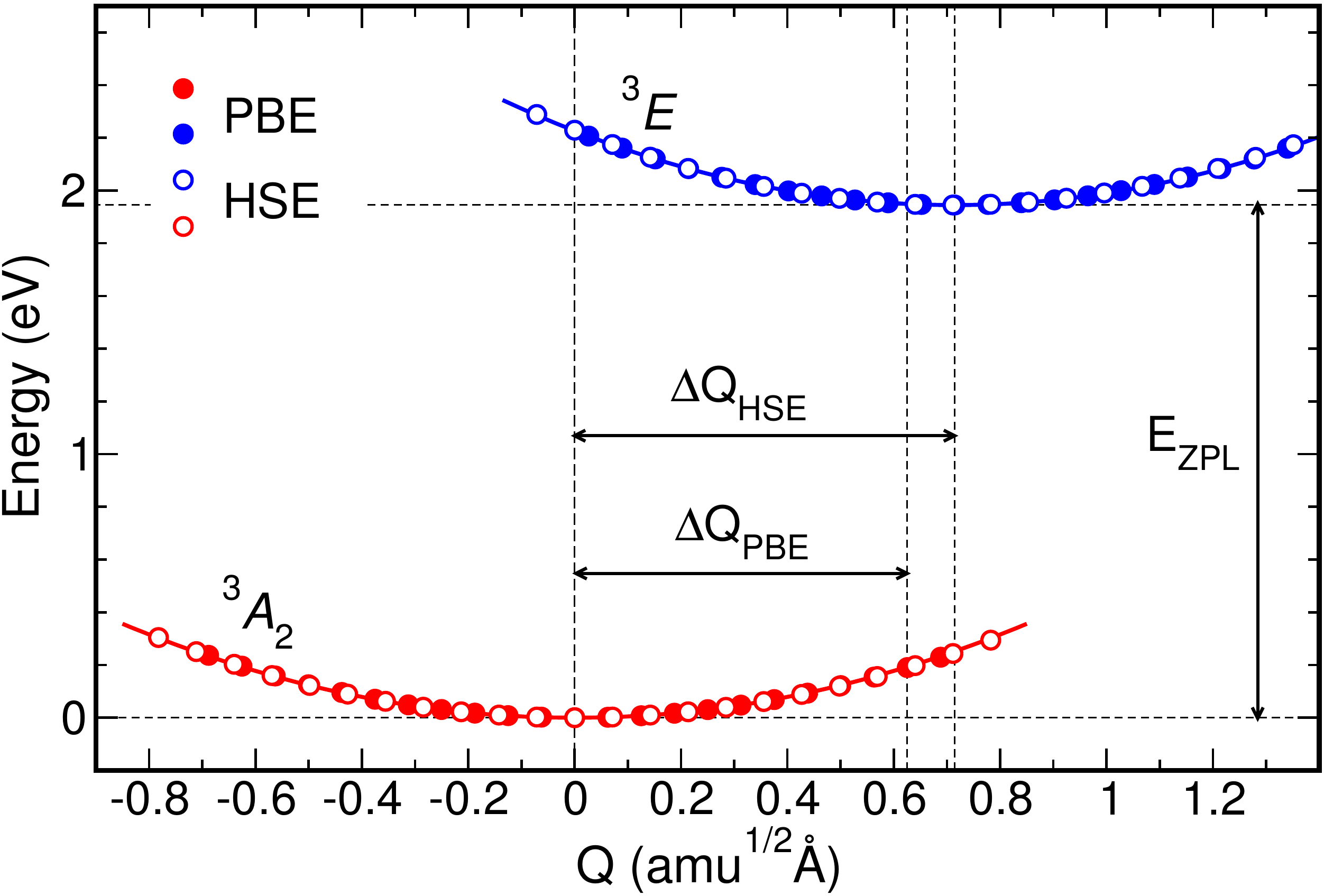}
\caption{1D cc diagram of the NV centre in the $^3\hspace{-0.5mm}A_2$ ground state and the $^3\hspace{-0.5mm}E$
excited state calculated using the PBE
and the HSE functionals. $E_{ZPL}=1.945$ eV is the zero-phonon line;
$\Delta Q$ (equation\ \ref{vib2}) is the mass-weighted atomic displacement between the minima in the ground and the
excited state. For a more meaningful comparison the $^3\hspace{-0.5mm}E$ potential energy curve as calculated in
PBE is shifted horizontally but $\Delta Q_{HSE}-\Delta Q_{PBE}$. The calculations used a $4\times 4\times 4$ (512-atom)
supercell.
}
\label{1D-CCD}
\end{center}
\end{figure}

In figure \ref{1D-CCD} we present an explicit calculation of the 1D cc diagram for the NV centre (results
from the $4\times4\times4$ supercell were used).
The HSE calculations (filled disks) yield $\Delta Q_{HSE}$=0.71 \AA$\cdot$amu$^{1/2}$ and $\Delta R_{HSE}$=0.20 \AA.
The PBE calculations (open disks) yield somewhat smaller values, $\Delta Q_{PBE}$=0.62 \AA$\cdot$amu$^{1/2}$ and
$\Delta R_{PBE}$=0.18 {\AA}.
(In passing, we note that in their seminal paper Davies and Hamer \cite{Davies_PRS_1976}
also estimated the total displacement $\Delta R$ based on a simple model for the defect.
Despite the fact that their model turned out to be not entirely correct, their
estimated $\Delta R=0.18$ \AA $ $ is astonishingly close to accurate first-principles results.)
In order to more meaningfully compare HSE and PBE results, we show the 1D cc diagram calculated
at the PBE level on the same graph, but shift the potential energy curve of the excited state
horizontally to $Q=\Delta Q_{HSE}$. Both the HSE and PBE curves are adjusted vertically to
match the experimental $E_{ZPL}=1.945$ eV, as discussed in Section \ref{experiment}.
A simple visual inspection of figure \ref{1D-CCD} then shows that if plotted this way the
potential energy curves determined in the two approaches lie virtually on top of each other.

More quantitatively, we have performed numerical fits to these one-dimensional potential energy curves
using the function $E(Q)=1/2\Omega^2 Q^2+\beta Q^3$. It can be easily shown that $\Omega$ determined in this
way is the mean square average of all the phonon modes contributing to the phonon sideband, the
weight of phonon mode $k$ being given by $q_k^2$ (equations\ (\ref{qk}) or (\ref{qk2})).
For the two functionals, these average frequencies differ by 1\% in the ground state, and 1.6\% in
the excited state, and in all cases the coefficient $\beta$ is essentially negligible.
These findings justify the assumptions at the beginning of this section.

The similarity of vibrational modes calculated in PBE and HSE can also be demonstrated by a direct
calculation of the vibrational spectrum of the supercell. Because of the high computational cost of
the HSE calculation, we have performed this calculation only for the smallest $2\times2\times2$
supercell. Vibrational modes and frequencies calculated using the two functionals
are indeed very similar, supporting the conclusion achieved by analysing figure \ref{1D-CCD}.
Therefore, it is a very good approximation to use vibrational modes calculated at the PBE level in all
calculations, and we adopt it for the present study.

The main difference between PBE and HSE functionals are the atomic relaxations
$\Delta Q$ (or $\vec{R}_{e} - \vec{R}_{g}$). It is because of this difference
that spectral functions of electron-phonon coupling $S(\hbar\omega)$ and
total Huang-Rhys factors (figure \ref{Sw-all}) are different in the two approaches.

\section{Calculations for very large supercells \label{supercell}}

A direct evaluation for the dilute limit, i.e., the use of very large supercells that would yield a
converged $S(\hbar\omega)$, is nearly impossible not only for an HSE hybrid functional, but also for
a less expensive PBE functional. This applies, in particular, to the calculation of vibrational modes.
To obtain results for large systems, we have used the following methodology.

For the two smallest supercells, i.e., $2$$\times$$2$$\times$$2$ (64 lattice sites) and
$3$$\times$$3$$\times$$3$ (216 sites) a direct approach has been applied. In particular, partial HR factors
$S_k$ have been evaluated using equations\ (\ref{Sk}) and (\ref{qk}). Vibrational modes
and frequencies have been determined by diagonalizing dynamical matrices obtained directly
from the supercell calculation.

\begin{figure}
\begin{center}
\includegraphics[width=13cm]{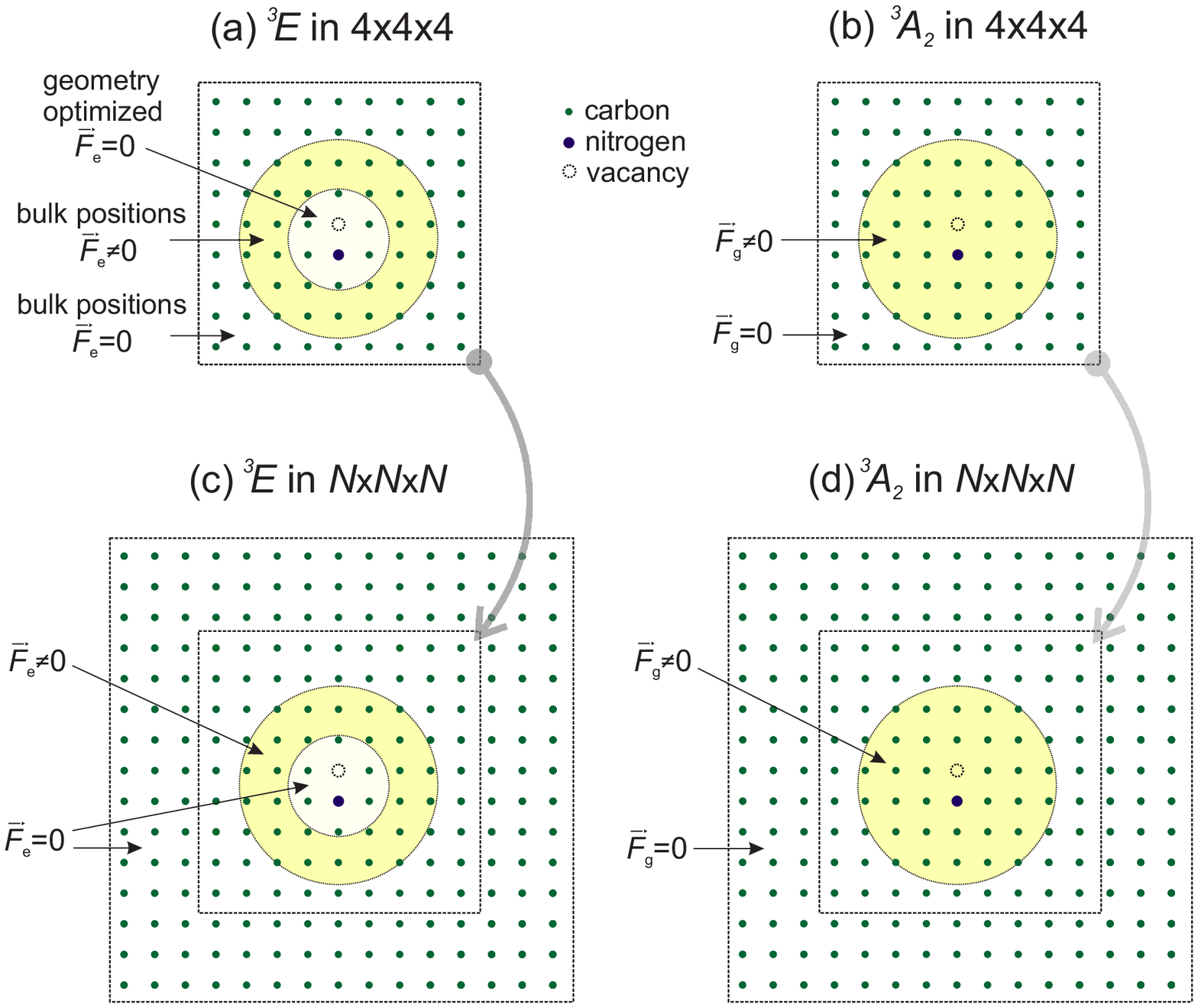}
\caption{The methodology to determine the spectral function $S(\hbar\omega)$ for large supercells.
(a) Constrained geometry optimization in the $4$$\times$$4$$\times$$4$ supercell
for the electronic excited state; (b) single-point calculation in the $4$$\times$$4$$\times$$4$ supercell
for the electronic ground state using the geometry obtained in (a); (c) and (d) embedding calculations obtained
in the previous two steps into a larger supercell $N$$\times$$N$$\times$$N$. For both the $4$$\times$$4$$\times$$4$
and the $N$$\times$$N$$\times$$N$ ($N>4$) vibrational modes and frequencies have been determined as described in the text.}
\label{large}
\end{center}
\end{figure}

For larger supercells we have used an alternative approach. First, we performed constrained geometry
relaxations for the $4$$\times$$4$$\times$$4$  supercell (512 lattice sites) with a defect in the middle of the supercell.
In the calculations for the excited state the atoms within 3 \AA $ $ of the vacancy were allowed to relax, while the remaining
atoms were kept in their ideal lattice positions (figure \ref{large}(a)). This procedure yields zero forces
$F_{e;\alpha i}$ within this chosen radius (white inner circle in figure \ref{large}(a)). The forces $F_{e;\alpha i}$
are non-zero for the atoms that were kept in their bulk positions. However, actual calculations indicate that the forces
are appreciable only within $\sim$7 \AA $ $ away from the vacancy (i.e., about 4 \AA $ $ away from the atoms that were allowed to
relax, indicated as an outer yellow circle in figure \ref{large}(a)). The crucial point is that there are no net forces
exerted on atoms that are at the boundary of this supercell. Subsequently, we kept the geometry of the
defect as optimized according to this procedure, but determined the forces $F_{g;\alpha i}$ on atoms when the electronic
state is changed to that of the ground state (figure \ref{large}(b)). The resulting forces are non-zero
in the entire region, shown as a yellow circle in figure \ref{large}(b), but essentially vanish beyond it.
These two calculations yield the difference $(F_{e;\alpha i}-F_{g; \alpha i})$ needed to determine partial HR
factors via equations\ (\ref{Sk}) and (\ref{qk2}). The fact that the forces are essentially zero beyond the
yellow circle in figure \ref{large} (a) and (b) does not mean that these atoms stay in their bulk position.
If the constraints were relieved, these atoms would move to find their equilibrium positions during a full geometry
optimization, since the movement of their neighbours during this optimization would result in a build-up of forces.
The point is that equation (\ref{qk2}) includes this automatically.

To determine the vibrational spectrum for this supercell we have made use of the fact that in
covalent semiconductors the dynamical matrix is short-ranged. For example, tests show that the inclusion of five
nearest-neighbour shells is sufficient to obtain a vibrational spectrum of bulk diamond. Thus, if the atoms in the
defect system are further away from each other than 4 \AA, we set the dynamical matrix element to 0. Otherwise, if
one of the atoms is within 2.5 \AA $ $ of the vacancy or the nitrogen atom, the matrix element is taken from the calculation
of the $3$$\times$$3$$\times3$ supercell. For other pairs we use bulk diamond values. The choice or parameters
leads to a converged vibrational spectrum of the $4$$\times$$4$$\times$$4$ defect supercell. A similar procedure
to construct the dynamical matrix was recently used for defects in GaN by Shi and Wang \cite{Shi_PRL_2012}. Partial
HR factors are then determined from equations\ (\ref{Sk}) and (\ref{qk2}).

For larger supercells $N$$\times$$N$$\times$$N$ ($N>4$) the procedure is as follows.
First, the two $4\times4\times4$ defect supercells from the previous steps were embedded into a larger
$N$$\times$$N$$\times$$N$ supercell for both the excited (figure \ref{large}(a)) and the ground state
(figure \ref{large}(d)). This automatically yields the force difference $(F_{e;\alpha i}-F_{g; \alpha i})$ for this larger
system. The wavefunction of the NV centre is very localized, and thus we might expect that the actual
calculation for a larger supercell, were it possible, would yield very similar force difference $(F_{e;\alpha i}-F_{g; \alpha i})$.
It is now clear why the formulation based on equation\ (\ref{qk2}) is hugely advantageous. When the atoms away from
the defect are kept fixed in ideal bulk positions during the geometry optimization in figure \ref{large}(a) this excludes
the elastic interaction between periodically repeated replicas of the defect, and eventually enables embedding this smaller
system into a large one. Vibrational spectra for these larger supercells have been determined in the same way as for the
$4$$\times$$4$$\times4$ supercell. Using these techniques we were able to study supercells as large as $11$$\times$$11$$\times$$11$
(10648 lattice sites). Our procedure procedure yields results of nearly the same quality as if explicit
first-principles calculations were performed for these large supercells.

\section{NV centre in the $^{13}$C diamond lattice \label{isotope}}

In all above discussions, we have considered the NV centre in natural diamond, with the atomic mass of carbon atoms
set to 12.0111 a.m.u. This is useful when comparing calculations to ensemble measurements, as done in the present
work. Our results apply to NV centres in $^{12}$C diamond as well. We have verified that
the frequency of the dominant phonon mode $\hbar\omega_0$=$65.0$ meV, the Huang-Rhys factor $S$=$3.67$,
and the Debye-Waller factor $w_{ZPL}$=$2.4$\% in $^{12}$C diamond are within 0.1 \% of the values in natural diamond.

\begin{table}
\caption{Comparison of calculated parameters pertaining to the phonon sideband in natural and $^{13}$C diamond.
$m_{\rm C}$ is the  mass of the carbon atom, $\hbar\omega$ is the energy of the most pronounced phonon mode, $S$ is the total
Huang-Rhys factors, and $w_{ZPL}$ is the weight of the zero-phonon line.}
\label{table_c_13}
\begin{indented}
\item[] \begin{tabular}{@{}c c c c c}
\br
                  & $m_{C}$ (a.m.u.)   &  $\hbar\omega_0$  &  $S$    & $w_{ZPL}$ \\
\mr
natural diamond/$^{12}$C diamond   &  12.0111 &  65.0           &  3.67   & 3.8 \%    \\
$^{13}$C diamond  &  13      &  63.4          &  3.80   & 3.4 \%    \\
\br
\end{tabular}
\end{indented}
\end{table}

It is of interest, however, to study NV centres
in $^{13}$C diamond, where the different mass of carbon atoms may lead to more
noticeable changes.
The comparison of the main parameters pertaining
to the vibrational sideband of NV centres in natural (or $^{12}$C) diamond vs. $^{13}$C diamond is shown in
Table \ref{table_c_13}. Calculations have been performed at the HSE level.
Compared to NV centres in natural diamond, the total HR factor $S$ increases by a factor of 1.035, i.e.,
slightly less than a factor $\sqrt{13/12.0111}\approx1.04$ which would be expected if only carbon atoms
would couple to the optical transition. This corresponds to a decrease of $w_{ZPL}$ from 3.8\% to 3.4\%.
The energy of the most pronounced phonon decreases only by a factor of 1.02, from 65.0 to 63.4 meV.

\vspace{5mm}


\end{document}